\documentclass[12pt]{article}

\usepackage{amsmath,amssymb}
\usepackage{bm}
\usepackage{cite}
\usepackage{comment}
\usepackage{sectsty}
\sectionfont{\large}

\let\OLDthebibliography\thebibliography
\renewcommand\thebibliography[1]{
  \OLDthebibliography{#1}
  \setlength{\parskip}{0pt}
  \setlength{\itemsep}{0pt plus 0.3ex}
}

\newcommand\x{\mathbf{x}}
\newcommand\p{\mathbf{p}}
\newcommand\q{\mathbf{q}}
\renewcommand\k{\mathbf{k}}
\newcommand\+{\dagger}
\newcommand\<{\langle}
\renewcommand\>{\rangle}
\renewcommand\d{\partial}
\renewcommand\L{\mathcal{L}}
\newcommand\Pf{\mathop{\mathrm{Pf}}}

\begin{document}

\title{\bf The Dirac Composite Fermion of the \\ Fractional Quantum Hall
  Effect}


\author{{\normalsize Dam Thanh Son}\\
{\small\em Kadanoff Center for Theoretical Physics, University of Chicago, Chicago, Illinois 60637, USA}}
\date{\empty}


\maketitle

\begin{abstract}
We review the recently proposed Dirac composite fermion theory of the
half-filled Landau level.
\end{abstract}

\section{INTRODUCTION}

The fractional quantum Hall effect (FQHE) was discovered in
1982~\cite{Tsui:1982yy}, only a couple of years following the
discovery of the integer quantum Hall effect (IQHE).  Being one of the most
nontrivial problems of condensed matter physics, the FQHE has
attracted the attention of theorists ever since.  One of the earliest and
most influential works is that by Laughlin~\cite{Laughlin:1983fy}.
The aim of the current review is to survey recent progress in the
understanding of one particular, but important, aspect of the FQHE:
the composite fermion (CF) in the half-filled Landau
level~\cite{Halperin:1992mh}.  In particular, we review the
arguments leading to the Dirac CF
theory~\cite{Son:2015xqa}.

The quantum Hall problem is attractive for theorists partly because of
its very simple starting point: a Hamiltonian describing particles
moving on a two-dimensional plane, in a constant magnetic field, and
interacting with each other through a two-body potential,
\begin{equation}\label{H-TOE}
  H = \sum_{a=1}^N \frac{(\p_a +  \mathbf{A}(\x_a))^2}{2m}
  + \sum_{\<a,b\>} V(|\x_a-\x_b|).
\end{equation}
Here $\mathbf{A}$ is the gauge potential corresponding to a constant
magnetic field.  The two-body potential $V$ is normally taken to be
the Coulomb potential $V(r)=e^2/r$, but many results
are valid for a large class of repulsive interactions.  The quantum
Hall states are characterized by many physical properties, including a
quantized Hall resistivity, a vanishing longitudinal resistivity, a
bulk energy gap, edge modes, etc.  For the purpose of this article, we
take the existence of an energy gap to be the defining property of the
quantum Hall states.  A very simplified summary of the experimental
situation is as follows: For certain values of the filling factor,
defined as
\begin{equation}
  \nu = \frac \rho{B/2\pi} ,
\end{equation}
where $\rho$ is the two-dimensional electron density, the system is in
one of the quantum Hall states with an energy gap.  The values of
$\nu$ for which there is a gap are either integers, in which case we
have an IQHE, or rational numbers, which correspond to the FQHE.

The existence of a gap for integer $\nu$ can be understood on the
basis of the approximation of noninteracting electrons.  In a
magnetic field $B$, the energy eigenvalues of the one-particle
Hamiltonian are organized into Landau levels,
\begin{equation}
 E_n = \frac Bm \left( n+\frac12 \right).
\end{equation}
The degeneracy of each Landau level is $B/2\pi$ per unit area.  At
integer $\nu$, states with $n<\nu$ are filled and those with $n\ge\nu$
are left empty.  The system then has a gap equal to the spacing
between Landau levels, which is $\omega_c=B/m$.

In contrast to the IQHE, the FQHE cannot be
understood from the noninteracting limit.  For example, when
$0<\nu<1$, the lowest Landau level (LLL; $n=0$) is partially filled,
so the noninteracting Hamiltonian has an exponentially large (in the
number of electrons) ground-state degeneracy.  The miracle of the FQHE
is that for certain rational values of $\nu$, interactions between
electrons lead to a gap.

There are two energy scales in the fractional quantum Hall (FQH) problem.  The first scale is
the cyclotron energy $\omega_c=B/m$, whereas the second scale is the
interaction energy scale.  In the case of the Coulomb interaction, the
latter energy scale can be estimated as the potential energy
between two neighboring electrons,
\begin{equation}
  \Delta = \frac{e^2}r \sim {e^2}{\sqrt B}.
\end{equation}
The FQH problem is usually considered in the limit
$\Delta\ll\omega_c$.  This limit is reached experimentally by taking
$B\to\infty$ at fixed $\nu$; theoretically it is also reached by
taking $m\to0$ at fixed $B$.  When $\Delta\ll\omega_c$ one can ignore
all Landau levels above the lowest one, and the problem can be
reformulated as pertaining to a Hamiltonian which operates only on the
LLL,
\begin{equation}\label{H-projected}
  H = \mathcal P_{\rm LLL} \sum_{\<a,b\>} V(|\x_a-\x_b|),
\end{equation}
where $P_{\rm LLL}$ is the projection to the LLL.
This extremely simple Hamiltonian, believed to underlie all the
richness of FQH physics, cannot be solved by traditional methods of
perturbation theory owing to the lack of a small parameter.  In
particular, there is only one energy scale---the Coulomb energy scale
$\Delta$.  The FQH problem is essentially nonperturbative.

\section{FLUX ATTACHMENT}

One of the most productive ideas in the FQH physics has been the idea
of the CF.  Theoretically, the notion of the CF
itself comes from another concept called flux
attachment~\cite{Arovas:1985yb}, which was applied to the FQHE in a
number of groundbreaking
works~\cite{Zhang:1988wy,Jain:1989tx,Fradkin:1991wy,Halperin:1992mh}.
Here, I review the standard textbook field theory of the CF,
although later on I argue that it needs some nontrivial
modification to become the correct low-energy effective theory.

In the FQH case, one ``attaches'' an even number (in the simplest
case, two) of magnetic flux quanta to an electron, transforming it to
a new object called the CF.  In the field theory
language, one starts from a theory of interacting electrons $\psi_e$
in (2+1) dimensions in a background magnetic field
\begin{equation}\label{L-orig}
  \L = i \psi_e^\+ (\d_t - i A_0) \psi_e 
  - \frac1{2m}|(\d_i-iA_i)\psi_e|^2 + \ldots
\end{equation}
where $\ldots$ stand for interaction terms, and derives, following
a certain formal procedure, a new Lagrangian for the CF
$\psi$,
\begin{equation}\label{L-cf}
  \L = i\psi^\+ (\d_t -iA_0 + ia_0)\psi
  - \frac1{2m} |(\d_i - iA_i + ia_i)\psi|^2 +
  \frac12 \frac1{4\pi} \epsilon^{\mu\nu\lambda}
   a_\mu \d_\nu a_\lambda + \ldots
\end{equation}
The Chern-Simons (CS) term in Eq.~(\ref{L-cf}) encodes the idea of flux
attachment.  In fact, the equation of motion obtained by
differentiating the action with respect to $a_0$ reads
\begin{equation}\label{flux_att}
  2 \psi^\+ \psi = \frac b{2\pi} ,
  \qquad b=\bm{\nabla}\times \mathbf{a},
\end{equation}
which means that the magnetic fluxes of the dynamic gauge field
$a_\mu$ are tied to the location of the CFs, with two
units of fluxes per particle.

There are two features of the field theory (\ref{L-cf}) (which I
call the HLR field theory after Halperin, Lee \& Read, who used it
to study the half-filled Landau level~\cite{Halperin:1992mh}) that
are rather trivial but worth listing here for future reference:
\begin{itemize}
\item The number of CFs is the same as the number of
  electrons.  It cannot be otherwise if the CF results
  from attaching magnetic fluxes to an electron.
\item The action contains a Chern-Simons term for $a_\mu$.  As
  demonstrated above, this term encodes in mathematical terms the idea
  of flux attachment.
\end{itemize}

In the literature, it is often stressed that transformation from
Eq.~(\ref{L-orig}) to Eq.~(\ref{L-cf}) can be done in an exact way (see, e.g.,
Ref.~\cite{Fradkin:1991wy}).  Unfortunately that also means that the
theory~(\ref{L-cf}) cannot be solved exactly.  To make any progress at
all, one must start with some approximation scheme, and in every
work so far this has been the mean field approximation in which one
replaces the dynamical gauge field $a_\mu$ by its average value
determined from Eq.~(\ref{flux_att}).  Because in the Lagrangian
(\ref{L-cf}) the gauge fields $A$ and $a$ enter through the difference
$A-a$, and the density of the CFs is the same as the
density of the original electrons, the effective average magnetic
field acting on $\psi$ is
\begin{equation}
  B_{\rm eff} = B - \<b\> = B-4\pi\rho.
\end{equation}
Translated to the language of the filling factors, 
\begin{equation}
  \nu = \frac\rho{B/2\pi}\quad \textrm{and} \quad \nu_{\rm CF} = \frac\rho{B_{\rm eff}/2\pi} ,
\end{equation}
the equation becomes
\begin{equation}
  \nu_{\rm CF}^{-1} = \nu^{-1}-2.
\end{equation}
In particular, the values $\nu=\frac{n}{2n+1}$ map to $\nu_{\rm
  CF}=n$.  In this way, we have mapped the FQH problem for the electrons
to the integer quantum Hall (IQH) problem for the CFs, which gives an
explanation for the emergence of an energy gap.  Experimentally,
one finds quite robust quantum Hall plateaus at these values of $\nu$,
up to $n\approx10$.

Another sequence of quantum Hall plateaus are found at
$\nu=\frac{n+1}{2n+1}$.  Now $\nu>\frac12$ so the effective average
magnetic field $B_{\rm eff}$ is negative; i.e., $B_{\rm eff}$ points to the
direction opposite to the direction of the original $B$.  The
CFs still form IQH states, with $n+1$ filled Landau
levels [$\nu_{\rm CF}=-(n+1)$].  Together, the two series of FQH
plateaus at $\nu=\frac n{2n+1}$ and $\nu=\frac{n+1}{2n+1}$ are called
the Jain sequences of plateaus.

One of the most spectacular successes of the CF theory
is the prediction of the nature of the $\nu=\frac12$ state (the
half-filled Landau level)~\cite{Halperin:1992mh}.  At this filling
fraction, the average effective magnetic field is equal 0, and the
CF should form a gapless Fermi surface.  HLR theory
thus predicts that the low-energy excitation is the fermionic
quasiparticle near the Fermi surface.  There is strong experimental
evidence that this is indeed the
case~\cite{Willett:1990,Kang:1993,Goldman:1994zz} and that the CF is a real
physical object---a quasiparticle near half filling---and not just a
mathematical construct.

Despite its astounding success, the quantum field theory~(\ref{L-cf})
has been criticized on various grounds.  The criticism leveled most
often against the theory~(\ref{L-cf}) is the lack of any information
about the projection to the LLL.  In particular, the
energy gap predicted by the mean-field picture is $B_{\rm eff}/m$,
which, for generic $\nu$, is of order $\omega_c$ instead of $\Delta$.  To
remedy the issue, one has to assume that the energy gap is determined
by an effective mass $m_*$, postulated to be parametrically
$B/\Delta$.  In particular, $m_*$ is assumed to remain finite in the
limit $m\to0$.

In fact, there are two parts to the energy scale problems.  The first
part is to derive, from microscopic calculations, the finite value of
$m_*$ in the limit $m\to0$.  This is a difficult problem and we
limit ourselves here by noting a few past attempts to address
it~\cite{Shankar:1997zz,Pasquier:1998sre,Read:1998dn}.  However, if
our goal is only to capture the low-energy physics, i.e., physics at
energy scales much smaller than the Fermi energy, then one should
simply take the effective mass $m_*$ as an input parameter, as in
Landau's Fermi liquid theory.  The second problem is to make the
low-energy effective field theory with $m_*$ consistent with the
fundamental symmetries of the original theory of electrons with a much
smaller mass.  Here one expects a relationship similar to the
relationship between the effective mass $m_*$ and the Fermi-liquid
parameter $F_1$ in Landau's Fermi liquid theory.  This program was
pursued in the 1990s and lead to the development of improvements
to the HLR theory like the ``magnetized modified RPA (MMRPA)'' of Simon,
Stern and Halperin.
(For a field-theoretic interpretation of the MMRPA, see
Ref.~\cite{Levin:2016wum}).  In principle, the problem can also be solved by using the
Newton-Cartan formalism, developed to make explicit the role
of Galilean symmetry (see, e.g.,
Refs.~\cite{Son:2013rqa,Jensen:2014aia,Geracie:2015xfa,Geracie:2015dea}).

Interestingly, most recent progress in the physics of the half-filled
Landau level has arrived from an attempt to address another problem,
historically regarded as less important and subordinate to the energy
scale problem: the lack of particle-hole (PH) symmetry.

\section{THE PROBLEM OF PARTICLE--HOLE SYMMETRY}

A system of nonrelativistic particles interacting through a two-body
interaction has two discrete symmetries: parity, or spatial reflection
($x\to x$, $y\to -y$), which we denote as $P$, and time reversal,
which we call $T$.  In a constant uniform magnetic field both
$P$ and $T$ are broken, but $PT$ is preserved.  $PT$ transforms the
wave function in the following way
\begin{equation}
  \Psi(x_i,y_i) \to \Psi'(x_i,y_i) =\Psi^*(x_i,-y_i).
\end{equation}
Interestingly, all wave functions that have been proposed, including
the Laughlin~\cite{Laughlin:1983fy} and Moore-Read~\cite{Moore:1991ks}
wave functions, are invariant under $PT$.

In the LLL limit ($\Delta\ll\omega_c$), the projected
Hamiltonian~(\ref{H-projected}) has an additional discrete symmetry:
the PH symmetry, first considered in
Ref.~\cite{Girvin:1984zz}.  To define the PH symmetry, one
chooses a particular basis of LLL one-particle states $\psi_k(x)$.
This basis defines the electron creation and annihilation operators
$c_k^\+$ and $c_k$.  The many-body LLL Fock space is obtained by acting
products of creation operators on the empty Landau level
$|\textrm{empty}\>$.  PH conjugation, $\Theta$, is defined
as an antilinear operator, which maps an empty Landau level to a full
one
\begin{equation}
  \Theta: |\textrm{empty}\> \to |\textrm{full}\> = \prod_{k=1}^M
  c_k^\+ |\textrm{empty}\>,
\end{equation}
where $M$ is the number of orbitals on the LLL.  It also maps a
creation operator to an annihilation operator and vice versa,
\begin{equation}
  \Theta: c_k^\+ \leftrightarrow c_k .
\end{equation}

One can show that the projected Hamiltonian maps to itself, up to an
addition of a chemical potential term,
\begin{equation}
  \Theta: H_{\rm LLL} \to H_{\rm LLL} - \mu_0 \sum_k c^\+_k c_k ,
\end{equation}
where $\mu_0$ depends on the interaction $V$.  This means that for
$\mu=\mu_0/2$, the Hamiltonian $H_{\rm LLL}-\mu N$ maps to itself: At
this chemical potential the Hamiltonian is PH symmetric.

Under PH conjugation the filling factor $\nu$ transforms as
\begin{equation}
  \nu \to 1-\nu .
\end{equation}
In particular $\nu=1/2$ maps to itself under PH conjugation: The
half-filled Landau level is at the same time half empty.  Furthermore,
$\nu=\frac n{2n+1}$ maps to $\nu=\frac{n+1}{2n+1}$: the two Jain
sequences of quantum Hall plateaus form pairs that map to each other
under PH conjugation: $\nu=1/3$ and $\nu=2/3$, $\nu=2/5$ and
$\nu=3/5$, etc.

\subsection{Status of Particle-Hole Symmetry in the Halperin--Lee--Read Theory}

Let us now ask what are the discrete symmetries of the HLR field
theory~(\ref{L-cf}).  It is easy to see that there is only one such
symmetry, $PT$.  The Chern-Simons theory does not have any discrete
symmetry which can be associated with PH conjugation.  This
reflects on the asymmetry in the treatment of quantum Hall plateaus:
the $\nu=\frac n{2n+1}$ is described by an integer quantum Hall state
where the CFs fill $n$ Landau levels, while its PH conjugate
$\nu=\frac{n+1}{2n+1}$ by $n+1$ filled Landau levels.

The Fermi liquid state with $\nu=1/2$ presents a particularly baffling
problem for PH symmetry.  Na\"{\i}vely, one expects PH
conjugation to map a filled state to an empty state and vice versa.
This would mean that the Fermi disk of the CFs, describing the Fermi
liquid state, maps to a hollow disk in momentum states: The states
with momentum $|\k|>k_F$ are filled, and those with $|\k|<k_F$ are
empty.  This is obviously silly.

What makes the PH symmetry problem seem hard is that PH symmetry is
not the symmetry of nonrelativistic electrons in a magnetic field [the
theory~(\ref{H-TOE})].  It only emerges as the symmetry after taking
the lowest Landau level limit [theory~(\ref{H-projected})].  The
PH symmetry of the LLL is not realized as a local operation
acting on fields.

Per se, the different treatments of the two Jain's sequences still
do not imply that the HLR theory is PH asymmetric.  It is
logically possible that, despite the appearance to the contrary, the
HLR theory would give physical results consistent with PH
symmetry.  This nontrivial theoretical possibility has been pursued
recently.  Wang et al.~\cite{Wang:2017cmz} computed the locations
of the magnetoroton minima in the Jain states with $\nu=\frac n{2n+1}$
and $\nu=\frac{n+1}{2n+1}$ at large $n$ using the HLR theory and found
that, surprisingly, these locations are symmetric to leading and
next-to-leading orders in $1/n$ (and coincide with the values
calculated~\cite{Golkar:2016thq} in the Dirac CF theory
described later in this review).

Historically, one of the earliest puzzles posed by PH
symmetry on the HLR theory has been recognized in 1997 by Kivelson et al.~\cite{Kivelson:1997}.  When disorders are
statistically PH symmetric, PH symmetry implies
that at half filling $\sigma_{xy}$ is exactly $\frac12(e^2/h)$.  In
the random phase approximations of the HLR theory, the resistivity
tensor of the electrons is directly related to the resistivity of the
CFs: $\rho_{xy}=\rho_{xy}^{\rm CF}+2h/e^2$.  The
CFs move in an average zero magnetic field, hence one
can set $\rho_{xy}^{\rm CF}=0$, which implies that
$\rho_{xy}=2(h/e^2)$.  These two results disagree with each other when
the longitudinal conductivity $\sigma_{xx}$ (or equivalently, the
longitudinal resistivity $\rho_{xx}$) is nonzero.

Here too, however, the more recent analysis by Wang et al.\
finds that the calculation of the Hall conductivity of the CFs may be more subtle than previously thought~\cite{Wang:2017cmz}.
The reason is that one can not treat the density of the CFs as a constant in the presence of disorder potential.  Detailed
calculations by Wang et al.\ show that, in certain regimes, the
CFs have Hall conductivity $\sigma_{xy}^{\rm CF}=-\frac12
e^2/h$, exactly the value required so that the Hall conductivity of
the electrons is $\frac12 e^2/h$ as dictated by PH
symmetry.

However, it seems that that the behavior of the quantities considered
in Ref.~\cite{Wang:2017cmz} are exceptions rather than the rule.  In
fact, many other physical observables do not behave in such a
fortuitous way.  For example, the susceptibility of the two Jain
states, $\nu=\frac{n}{2n+1}$ and $\nu=\frac{n+1}{2n+1}$, computed from
the MMRPA, differ from each other by a factor of
$(n+1)^2/n^2$~\cite{Nguyen:2017mcw}.  Another example of the violation of
PH symmetry in the HLR theory was found in
Ref.~\cite{Levin:2016wum}.  In that study, it was shown that PH
symmetry strictly determines the first $q^2$ correction to the Hall
conductivity in the regime $\omega\gg v_F q$,
\begin{equation}
  \sigma_{xy}(\omega, q) = \frac12 \left[1-\frac14 (q\ell_B)^2\right]
\end{equation}
but the MMRPA of the HLR theory fails to get the correct value of
$-\frac14$ for the $q^2$ correction; instead, one would get 0
following the prescription of MMRPA.  The discrepancy was traced to
the incorrect value of orbital spin assigned to the CF
by the flux attachment procedure.  One must keep in mind, however,
that the calculations in the HLR theory have been performed within a
random-phase approximation.  The question about the effect of
gauge-field fluctuations need to be further
investigated~\cite{You:2016xmw}.

\subsection{Spontaneous Breaking of Particle-Hole Symmetry?}

Another logical possibility is that the Fermi-liquid ground state of
the half-filled Landau level spontaneously breaks PH
symmetry.  This possibility was investigated by Barkeshli et
al.~\cite{Barkeshli:2015afa}.  If that is the case, there are two
states at $\nu=1/2$: One corresponds to a Fermi surface of ``composite
particles'' and the other to that of ``composite holes''; the two states are
degenerate with each other in the lowest-Landau level limit.  In fact,
it is believed that on the second Landau level, the ground state is
either a Pfaffian~\cite{Moore:1991ks} or an anti-Pfaffian
state~\cite{Levin:2007,SSLee:2007}, which are PH conjugates
of each other but are distinct from each other, breaking PH
symmetry spontaneously.

However, there is no numerical evidence for this kind of spontaneous
PH symmetry breaking in the Fermi liquid state.  In fact,
the experimental result of Ref.~\cite{Baldwin:2014} indicates,
at least na\"{\i}vely, that the $\nu=1/2$ Fermi liquid is equally well
interpreted as being made out of composite particles or
composite holes.  There is now strong numerical evidence that the
$\nu=1/2$ state is PH symmetric~\cite{Geraedts:2015pva}.

\section{DIRAC COMPOSITE FERMION}

Having argued against a hidden PH symmetry of the HLR
theory and a spontaneous breaking of PH symmetry, we now
consider the third, most nontrivial possibility: The low-energy
physics of the half-filled Landau level is described by a theory
different from the HLR theory.  The theory must satisfy PH
symmetry but also preserve all successful phenomenological predictions
of the HLR theory.  The Dirac composite fermion theory, proposed in
Ref.~\cite{Son:2015xqa}, satisfies these requirements.  The essence of
the theory is that the CF does not transform into a
composite hole under PH symmetry, but remains a
composite particle.  Only the momentum of the CF flips
sign under PH conjugation,
\begin{equation}\label{PH_CF}
  \Theta: \k \to -\k.
\end{equation}
Implicitly, we assume that the Fermi disk of the CFs
transform into itself (a filled disk, not a hollow disk).

Equation~(\ref{PH_CF}) is usually associated with time reversal.  In
the theory of Dirac CFs, the CF is described by a
two-component spinor field $\psi$, which transforms under PH
conjugation following the formula that is usually identified with
time reversal,
\begin{equation}
 \psi \to i\sigma_2\psi .
\end{equation}

There are several arguments one can put forward to argue that the
CF has to be a massless Dirac particle.  One hint comes
from the CF interpretation of the Jain-sequence states.  Recall that
one problem with the standard CF picture is that the $\nu=\frac
n{2n+1}$ corresponds to the composite fermion filling factor $\nu_{\rm
  CF}=n$, whereas $\nu=\frac{n+1}{2n+1}$ maps to $\nu_{\rm CF}=n+1$
(ignoring the sign).  By constrast, these two states are
PH-conjugate pairs and should be described by the same filling factor of
the CF in any PH-symmetric theory.  The most na\"{\i}ve way
to reconcile these different pictures is to replace the filling
factors $\nu_{\rm CF}=n$ and $\nu_{\rm CF}=n+1$ with the average value
$\nu_{\rm CF}=n+\frac12$.  But now we have a problem: We want to map
the FQHE in the Jain sequences to the IQHE of the CFs,
but is it possible to have an IQH state with half-integer filling
factor?  Indeed, it is, if the composite fermion is a massless Dirac
fermion.  Half-integer quantization of the Hall conductivity is a
characteristic feature of the Dirac fermion, confirmed in experiments
with graphene~\cite{Novoselov:2005kj,Zhang:2005zz}.

The second argument in favor of the Dirac nature of the CF relies on a
property of the square of the PH conjugation operator
$\Theta^2$~\cite{LevinSon,Geraedts:2015pva}.  It is intuitively clear that
applying PH conjugation twice maps a given state to itself,
but there is a nontrivial factor of $\pm1$ that one gains by doing so.
Consider a generic state on the LLL with $N_e$ electrons,
\begin{equation}
  |\psi\> = \prod_{i=1}^{N_e} c_{k_i}^\+ \, |\textrm{empty}\>.
\end{equation}
Then under PH conjugation
\begin{equation}
  \Theta: |\psi\> \to \prod_{i=1}^{N_e} c_{k_i}^{\phantom{\dagger}} |\textrm{full}\>
  = \prod_{i=1}^{N_e} c_{k_i}^{\phantom{\dagger}} 
    \prod_{j=1}^{M} c_j^\+ |\textrm{empty}\>.
\end{equation}
Applying $\Theta$ again one finds
\begin{equation}
  \Theta^2: |\psi\> \to \prod_{i=1}^{N_e} c_{k_i}^\+ \prod_{j=1}^M c_j
  |\textrm{full}\> =
   \prod_{i=1}^{N_e} c_{k_i}^\+ \prod_{j=1}^M c_j^{\phantom{\dagger}}
  \prod_{k=1}^M c_k^\+ |\textrm{empty}\> = (-1)^{M(M-1)/2}|\psi\>.
\end{equation}
This relationship is quite easy to interpret when $M$ is an even
number: $M=2N_{\rm CF}$.  Then
\begin{equation}\label{Theta2NCF}
  \Theta^2: |\psi\> \to (-1)^{N_{\rm CF}} |\psi\>.
\end{equation}
This formula suggests the following interpretation: $N_{\rm CF}$ is
the number of CFs of the state $|\psi\>$, and each
CF is associated with a factor of $-1$ under
$\Theta^2$.  This $-1$ factor is natural for Dirac fermion.

In order to have a correct $\Theta^2$, we have to identify the number
of CFs with half the number of orbitals on the LLL:
$N_{\rm CF}=M/2$, which is \emph{independent} of the number of
electrons $N_e$.  This contradicts the intuitive picture of flux
attachment, in which the CF is obtained by attaching
two units of flux quanta to an electron.  However, that is
expected: In a theory that treats particles and holes in a symmetric
way, the number of CFs has to be in general different
from the number of electrons, otherwise it would have to be equal to
the number of holes as well.

\subsection{The Field Theory of the Dirac Composite Fermion}

The minimal version of the theory of the Dirac CF has
the following Lagrangian,
\begin{equation}\label{L-dual}
  \L = 
  i\psi^\+ [\d_t -ia_0 - v_{\rm F}\sigma^i(\d_i - ia_i)]\psi 
  - \frac1{4\pi}
  \epsilon^{\mu\nu\lambda}A_\mu\d_\nu a_\lambda
  + \frac1{8\pi} \epsilon^{\mu\nu\lambda}A_\mu \d_\nu A_\lambda,
\end{equation}
where $a_\mu$ is a dynamical gauge field.  The kinetic term in
Eq.~(\ref{L-dual}) contains, instead of the speed of light, a Fermi
velocity $v_{\rm F}$, determined by microscopic physics.  The
Lagrangian~(\ref{L-dual}) is similar to (\ref{L-cf}) after the
replacement $a\to a+A$, with two differences.  One is the Dirac nature
of the CF $\psi$.  The other is the absence of the
CS term $ada$ in the Lagrangian: Such a term (as also the
mass term for $\psi$), if present, would disallow any discrete
symmetry that could be identified with PH symmetry.
Interestingly, each of such modifications to the HLR theory would
shift the filling factors of the Jain-sequence plateaus, but together
the shifts cancel each other, and the Jain sequences remain unchanged
(see below).

Differentiating~(\ref{L-dual}) with respect to $A_0$, one obtains the
electron density
\begin{equation}\label{rhob}
  \rho = \frac{\delta S}{\delta A_0} = \frac{B-b}{4\pi} \,.
\end{equation}
On the other hand, the equation of motion obtained by differentiating
the action with respect to $a_0$ is
\begin{equation}\label{rhoCFB}
   \bar\psi \gamma^0 \psi = \frac B{4\pi}\,,
\end{equation}
i.e., the CF density is set by the external magnetic field.  Thus, the
role of the magnetic field and the density is flipped when one goes
from the original electrons to the CFs.  This is the
salient feature of particle-vortex duality, well known for
bosons~\cite{Peskin:1977kp,Dasgupta:1981zz} but not, until this
example, for fermions.

If one defines the filling factors of the electron and the CF as
\begin{equation}
  \nu = \frac{2\pi\rho} B \quad \textrm{and} \quad \nu_{\rm CF} = \frac{2\pi\rho_{\rm CF}}b\,,
\end{equation}
respectively, then from Eqs.~(\ref{rhob}) and (\ref{rhoCFB}), we find that they are
related by
\begin{equation}
  \nu_{\rm CF} = - \frac1{4(\nu-\frac12)} .
\end{equation}
In particular, $\nu=\frac n{2n+1}$ maps to $\nu_{\rm CF}=n+\frac12$,
which is the filling factor of an integer quantum Hall state of a
Dirac fermion.  The PH conjugated state
$\nu=\frac{n+1}{2n+1}$ maps to $\nu_{\rm CF}=-(n+\frac12)$, which is
the same filling factor, but in a magnetic field of the opposite sign,
manifesting the PH symmetry.

It should be emphasized that the Dirac nature of the CF does not mean
that there is a Dirac cone for the CF.  The tip of the cone is at
$\k=0$, whereas the CF, as a low-energy mode, exists only near the Fermi
surface.  The Dirac nature of the CF, strictly speaking, only means
that the fermionic quasiparticle has a Berry phase of $\pi$ around the
Fermi surface.  It is easy to show that such a Berry phase follows
from Eqs.~(\ref{PH_CF}) and (\ref{Theta2NCF}).  The quasiparticle
Berry phase has been identified as a important ingredient of Fermi
liquids~\cite{Haldane:2004zz}, but the possibility of such a phase for
the CF in FQHE has been overlooked in the literature
until very recently.

\subsection{Galilean Invariance and Electric Dipole Moment}

It has to be emphasized that the energy scale problem still
exists, under a different guise, in the Dirac CF
theory.  This problem now appears as the problem of the origin of the
Fermi velocity of the CF.  The scale of the Fermi
velocity should be set by the interactions, but, as before, there is
no simple way to calculate it from the microscopic theory.

If one takes the Fermi velocity of the CF as an input
parameter, one still has to make sure that the theory of the CF exhibits the symmetry of the original theory, in particular,
the Galilean invariance.  It is crucial that Galilean invariance is
implemented into the effective theory; important properties, like the
$q^4$ behavior of the susceptibility at small $q$, follow from it.  However, the Dirac action is not invariant under Galilean
transformation,
\begin{equation}
  \psi(t,\x) \to \psi'(t,\x) = \psi(t,\x-\mathbf{v}t).
\end{equation}
To make the Dirac action Galilean invariant, one needs to add
an additional term to the effective Lagrangian, replacing
\begin{equation}\label{replacement}
  i \psi^\+ \d_t \psi \to i  \psi^\+\d_t \psi
  + \frac i2 \frac{\epsilon^{ij}E_j}B (\psi^\+ D_j\psi - D_j\psi^\+\psi).
\end{equation}
Because the electric field transforms under Galilean transformation as
$E_i\to E_i - \epsilon^{ij}v_j B$, the action is now Galilean
invariant.  The new term has a very intuitive physical meaning: It
implies that the CF has an electric dipole moment with
respect to the external electromagnetic field, with the dipole moment
perpendicular to the momentum:
$\mathbf{d}=-\ell_B^2\mathbf{p}\times\mathbf{\hat z}$, realizing an old idea
by Read~\cite{Read-neutral1,Read-neutral2}.

The dipole picture has been revived recently by Wang \& Senthil in
the context of the Dirac CF \cite{Wang:2016fql}.  They
suggest that the dipole moment provides a way to understand the origin
of the Berry phase of the CF around the Fermi surface.
In this picture, the CF is a pair made up of an electron and a
correlation hole that form an electric dipole, with the value of the
dipole moment perpendicular to the direction of the momentum.  When a
CF is dragged around the Fermi surface, the dipole
rotates and the fermion acquires a Berry phase because of the external
magnetic field.  A simple calculation shows that the Berry phase is
equal to $\pi$ when the CF makes a circle of radius
exactly equal to the Fermi momentum in momentum space.

The simplicity of this explanation of the Berry phase may be
deceptive, however.  If one takes the above dipole picture literally,
one finds not only a global Berry phase but also a local Berry
curvature that is uniform in momentum space.  A massless Dirac fermion, in
contrast, has only a global Berry phase but no local Berry curvature.
Thus, the connection between the Berry phase and the dipole moment may
not be straightforward.  Within the low-energy effective theory
embodied by Eqs.~(\ref{L-dual}) and (\ref{replacement}), the Berry
phase and the dipole moment appear as separate ingredients.

\section{CONSEQUENCES OF THE DIRAC COMPOSITE FERMION}

The Dirac CF theory has distinct consequences, which are, in
principle, verifiable in experiments and numerical simulations.  By
construction, it leads to PH-symmetric results where the
HLR theory violates PH symmetry.  For example, the
susceptibilities of the $\nu=\frac n{2n+1}$ and $\nu=\frac{n+1}{2n+1}$
states are now the same~\cite{Nguyen:2017mcw}.

It is numerical simulations~\cite{Geraedts:2015pva} that provide the
currently most nontrivial test of the Dirac nature of the CF.  The numerical finding is the disappearance, attributable to
particle-hole symmetry, of the leading $2k_F$ singularity in certain
correlation functions.  In a Fermi liquid, typical two-point
correlation functions exhibit nonanaliticity at twice the Fermi
momentum ($2k_F$ singularities) due to intermediate states involving a
particle and a hole near diametrically opposite points on the Fermi
surface.  In the case of the (2+1)d massless Dirac fermion, it is well
known that two-point correlation functions of time-reversal invariant
operators are free from the leading $2k_F$ singularity in a generic
two-point correlator.  This is caused by the quasiparticle Berry
phase $\pi$ around the Fermi surface that forbids the creation, by a
time-reversal invariant operator, of a PH pair where the particle and the
hole carry opposite momenta.  In the half-filled Landau level, the
role of time reversal is played by PH symmetry; therefore
to test the Berry phase one should look for the absence of the leading
$2k_F$ singularity in correlation functions of a PH symmetric operator.
The electron density operator $\rho=\psi_e^\dagger\psi_e$ is not PH
symmetric (the deviation of the density from the mean density,
$\delta\rho=\rho-\rho_0$, flips sign under PH conjugation), but one can
easily write down more complicated operators that are PH symmetric,
for example, $\delta\rho \nabla^2 \rho$.  In
Ref.~\cite{Geraedts:2015pva} the leading $2k_F$ singularity in the
correlation function of such an operator was shown to disappear when
PH symmetry is made exact (and to reappear when PH symmetry is
violated), confirming the Dirac nature of the CF.

There are also predictions about transport that are, strictly
speaking, consequences of PH symmetry.  If one introduces
the conductivities $\sigma_{xx}$ and $\sigma_{xy}$, and the
thermoelectric coefficients $\alpha_{xx}$ and $\alpha_{xy}$, so that
\begin{equation}
  \mathbf{j} = \sigma_{xx} \mathbf{E} + \sigma_{xy}\mathbf{E}\times
  \mathbf{\hat{z}} + \alpha_{xx} \bm{\nabla} T
  + \alpha_{xy} \bm{\nabla} T \times \mathbf{\hat{z}},
\end{equation}
then at exact half filling, PH symmetry
implies~\cite{Son:2015xqa,Potter:2015cdn}
\begin{equation}
  \sigma_{xy} = \frac12 \frac{e^2}h \quad \textrm{and} \quad \alpha_{xx}=0.
\end{equation}
A manifestly PH symmetric theory like the Dirac composite
fermion theory reproduces these results automatically~\cite{Potter:2015cdn}.

In real samples one expects that PH symmetry is not exact
due to the mixing of higher Landau levels.  When the breaking of
PH symmetry is small, one can expect its effect to be
parametrized by a single number: the Berry phase that the CF acquires along the Fermi circle~\cite{Potter:2015cdn}.

\subsection{The PH-Pfaffian State}

One consequence of the Dirac CF theory is the existence
of a new gapped state at half filling that is PH
symmetric.  When the CFs undergo Bardeen-Cooper-Schrieffer (BCS) pairing, the
resulting state is a gapped quantum Hall state.  In the traditional
HLR theory, the pairing between fermions has to be in a channel with
odd orbital moment to be consistent with Fermi statistics.  The Dirac
nature of the composite fermion, in contrast, requires the BCS gap to
carry an even orbital moment.  The simplest pairing channel
is~\cite{Son:2015xqa}
\begin{equation}\label{PH-Pfaffian}
  \< \varepsilon^{\alpha\beta}\psi_\alpha\psi_\beta \>,
\end{equation}
and does not break either $PT$ or PH symmetry.  The state
is the quantum Hall analog of the so-called T-Pfaffian state, a
proposed fully gapped state on the interacting surface of topological
insulators~\cite{Wang:2013uky,Bonderson:2013pla,Chen:2013jha,Metlitski:2015bpa}.
The PH-Pfaffian state is distinct from the Pfaffian state and the
anti-Pfaffian state, which are not PH symmetric and are the
PH conjugates of each others.  In particular, the shift of
the PH-Pfaffian state is equal to 1, whereas the shift of the Pfaffian is
3 and the anti-Pfaffian state carries shift of $-1$.

Within the Dirac composite fermion theory, the Pfaffian and
anti-Pfaffian states involve pairing with orbital moments $\pm2$,
i.e., with order parameters $\<
\varepsilon^{\alpha\beta}\psi_\alpha(\d_x\pm i\d_y)^2\psi_\beta \>$.
The Dirac CF thus provides a symmetric treatment of the
Pfaffian and anti-Pfaffian states, with the PH-Pfaffian state
situating exactly at the midpoint in terms of the orbital moment of
the Cooper pair.

In the HLR theory, the orbital moment of the Cooper pair is shifted by
one unit compared to the Dirac theory.  The Pfaffian state appears as
a $p_x+ip_y$ pairing of the CFs.  This interpretation
of the Pfaffian state has been known for a long time.  The PH-Pfaffian
state also corresponds to $p$-wave pairing, but here  the orbital moment is of the
opposite sign, i.e., $p_x-ip_y$ pairing.  Finally, to get the
anti-Pfaffian state from the usual HLR theory, one would pair the
CFs in the $\ell=-3$ orbital moment channel.  The HLR theory
does not explain why the $\ell=1$ and $\ell=-3$ paired states have the
same energy.

At this moment, there is no simple particle-hole symmetric
wavefunction for the PH-Pfaffian state (except ones that involve
explicit symmetrization of two PH conjugate wavefunctions).
However, there seem to exist wavefunctions that are quite close to
being PH symmetric.  One simple proposed wave function
is~\cite{Yang:2017} (omitting the Gaussian factor)
\begin{equation}
  \Psi(z_1, \cdots, z_n) =
  \Pf\left( \frac1{\d_{z_i}-\d_{z_j}} \right)
  \prod_{\<ij\>}(\d_{z_i}-\d_{z_j}) \prod_{\<ij\>} (z_i-z_j)^3.
\end{equation}
This wavefunction has a large overlap with its PH
conjugate, at least for relatively small number of particles.

It is not at all clear if the PH-Pfaffian state can be realized with
any two-body interactions.  Numerical simulations either show a Fermi
liquid state or a PH-breaking gapped
state~\cite{Geraedts:2015pva}.  On the other hand,
Ref.~\cite{Zucker:2016} argue that the PH-Pfaffian state may be
stabilized by the mixing with higher Landau levels and impurities (note
that the topological order of the PH-Pfaffian does not require
PH symmetry to exist) and it may present a viable
alternative for the $\nu=5/2$ plateau.

\section{UNDERSTANDING THE ORIGIN OF THE DIRAC\\ COMPOSITE FERMION}

As described above, much theoretical and numerical evidence
implies that the quasiparticle of the $\nu=1/2$ state has to be a
Dirac CF.  This means, in particular, that the old,
intuitive picture of a CF as an electron with two
attached flux quanta (or rather, an electron as a CF
with two attached flux quanta) cannot be literally correct.  

It is perhaps fair to say that at this moment there is no completely
satisfactory derivation of the Dirac CF theory from the
microscopic theory
of fermions on the LLL.
An attempt to embed the
Girvin--MacDonald--Platzman algebra into that of Dirac CF
was made in Ref.~\cite{Murthy:2016jnc}.  The role of the gauge field,
however, is not clear in this mapping.

Most other attempts to derive the theory of Dirac CFs
are actually aimed at deriving a zero magnetic field version: the
duality between the free Dirac fermion, defined by the action
\begin{equation}\label{LA}
  \mathcal L_A = i\bar\Psi\gamma^\mu (\d_\mu -i A_\mu) \Psi ,
\end{equation}
and a theory of a fermion coupled to an gauge field which we call
parity-invariant QED$_3$, or QED$_3$ in short,
\begin{equation}\label{LB}
  \mathcal L_B = i\bar\psi\gamma^\mu(\d_\mu - ia_\mu)\psi
  - \frac1{4\pi}\epsilon^{\mu\nu\lambda}A_\mu\d_\nu a_\lambda .
\end{equation}
(A more accurate version of $\mathcal L_B$ will be given later.)  It
is usually claimed that the duality between the two theories
(\ref{LA}) and (\ref{LB}) implies the CF theory of the
half-filled Landau level.  The argument runs as follows: A finite
magnetic field in free Dirac fermion theory corresponds to a finite
density in QED$_3$, and at finite density, the fermions of QED$_3$
form a Fermi liquid.  This line of argument, however, hides an
important subtle point.  For a free fermion in an external magnetic
field, the ground state is exponentially degenerate.  If the
conjectured duality is valid, then QED$_3$ must have finite entropy at
finite density---a very unusual situation in quantum field theory
indeed, which does not have a direct relationship with the quantum
Hall state at $\nu=1/2$.  What is implicitly assumed here is that
small interactions in theory A make the ground state in theory B
unique and Fermi-liquid like.  In particular, the Fermi velocity of
the CF is determined by terms not present in the basic
Lagrangians~(\ref{LA}) and (\ref{LB}).  One recognizes here the energy
scale problem in a different guise.

Another subtlety of the duality is the stability of the theory
(\ref{LB}).  The question is related to the question about the fate of
scale invariance in a theory of $N_f$ flavors of fermions coupled to a
gauge field.  At large $N_f$ the theory is conformal; however, it is
not clear what is the situation at small $N_f$.  Approaches based on
the gap equation (unreliable because they rely on uncontrolled
truncations of the Schwinger--Dyson equation) give rather high values
of $N_f^{\rm crit}$; however a lattice simulation~\cite{Karthik:2015sgq} implies that the
theory remains conformal down to $N_f=2$.

One promising approach to derive the duality is the ``wire
construction'' by Alicea, Mross \& Motrunich~\cite{Mross:2015idy}.
Here one discretizes the Dirac fermion theory into a set of
(1+1)-dimensional fermions and then applies the well-developed machinery
of bosonization to these theories.  One then applies a linear nonlocal
transformations on the bosons and refermionizes the resulting theory.
The result is a discretized version of the CF coupled
to a gauge field.  This approach provides the most explicit mapping of
fields between the two sides of the duality.  The existence of the
continuum limit is assumed, and the approach breaks rotational
invariance and continuous translational invariance.

In the rest of the section we will describe a promising approach
pioneered in recent works~\cite{Karch:2016sxi,Seiberg:2016gmd}.  This
approach is particularly interesting since it places the quantum Hall
problem in the context of a large ``web of dualities'' between
different (2+1)-dimensional gauge theories.


The duality between free Dirac fermion and parity-invariant QED$_3$ is
similar to the more familiar duality between the a complex scalar
field theory near the Wilson--Fisher fixed point (Wilson--Fisher
scalar) and that of a scalar field coupled to a U(1) gauge field
near the phase transition between the Coulomb phase and the Higgs
phase.  One of the most important recent insights has been the
understanding that the two dualities are consequences of a perhaps
more elementary, seed duality.  The latter duality is a relativistic
version of flux attachment.

In this section, the following shortcut notation is used: $ada\equiv
\epsilon^{\mu\nu\lambda}a_\mu\d_\nu a_\lambda$, $Ada\equiv
\epsilon^{\mu\nu\lambda}A_\mu\d_\nu a_\lambda$.  Moreover, capital
letters ($A$) refer to background gauge fields, lower-case letters
($a$, $b$, etc.) refer to dynamical gauge fields.  The seed duality is
between a relativistic bosonic theory coupled to a CS gauge
field $a$, and a relativistic fermionic theory,
\begin{equation}\label{boson_fa}
  \L (\phi, a) + \frac1{4\pi} ada + \frac1{2\pi}Ada 
  \Leftrightarrow 
  \L(\psi,A) - \frac12 \frac1{4\pi} AdA .
\end{equation}
Here $\L(\phi,a)$ includes all terms consistent with symmetry,
including the mass term $m^2|\phi|^2$ whose coefficient needs to be
tuned to a critical point,
and a $|\phi|^4$ term whose coefficient is assumed to flow, in the infrared,
to a Wilson-Fisher-like fixed point.
On the right-hand side is the theory
$\L(\psi,A)$, which involves a two-component fermion coupled to the external
field $A$.  Without the CS term $AdA$, the right-hand side is parity
and time-reversal invariant when the mass of the fermion is zero.  The
statement of the duality is that if one takes the path integral of the
exponent of the actions on both side over all fields except the
background fields (i.e., $\phi$ and $a$ on the left-hand side and
$\psi$ on the right-hand side), one gets the same functional of the
external field $A$.

The duality has not been proven but can be checked in the massive
limits.  In particular
\begin{itemize}
\item When the $\phi$ field is massive (with positive $m^2$), one can
  integrate out $\phi$ to get nothing.  The field $a$ can now be
  integrated out, and one gets, on the bosonic side, $-\frac1{4\pi}AdA$.
  On the fermionic side, this corresponds to a positive sign of the
  mass of the  fermion. Integrate out the fermion, one also gets
  $-\frac1{4\pi}AdA$.
\item When $\phi$ condenses (negative $m^2$), $a$ becomes massive.
  Integrating out $a$ now gives nothing.  On the fermionic side, the
  fermion mass is negative.  Integrating out the fermion, then one cancels
  out the $AdA$ CS term.
\end{itemize}

From the seed duality whole web of different dualities can be derived
in an almost mechanical manner.  We present here a few illustrative
cases.  If one gauges the gauge field $A$, replacing $A\to b$, and
added a term $\frac1{2\pi}Adb$, with $A$ being a new background field
and $b$ a field to be integrated over, one gets the following duality:
\begin{equation}
  \L(\phi, a) + \frac1{4\pi}ada + \frac1{2\pi}bda + \frac1{2\pi}Adb
  \Leftrightarrow
  \L(\psi, b) - \frac12\frac1{4\pi} bdb + \frac1{2\pi}Adb .
\end{equation}
On the left-hand side, one can integrate over $b$ and then $a$ to
obtain a new duality:
\begin{equation}\label{fermion_fa}
  \L(\phi, A) + \frac1{4\pi}AdA \Leftrightarrow
  \L(\psi, a) - \frac12\frac1{4\pi}ada + \frac1{2\pi}Ada .
\end{equation}
This can be interpreted as the statement that a fermion with attached
flux is a boson.  Note that the right-hand side is gauge
invariant: the CS term with half-integer coefficient can
be thought of as coming from the Pauli--Villars regularization of
$\L(\psi,a)$ when the Pauli--Villars mass goes to infinity.

The bosonic particle-vortex duality can now be derived from the
boson-fermion dualities by a series of formal manipulations.
We start from the duality~(\ref{fermion_fa}), move the $AdA$
term from the left-hand side to the right-hand side, and promote $A$
to a dynamical gauge field as done before.  We find 
\begin{equation}
  \L(\phi, b) + \frac1{2\pi} Adb \Leftrightarrow
  \L(\psi, a) - \frac12\frac1{4\pi}ada + \frac1{2\pi}bda 
  - \frac1{4\pi}bdb + \frac1{2\pi}Adb .
\end{equation}
The integral over $b$ on the right-hand side is a Gaussian integral,
with the saddle point at $b=a+A$.  We find after the integration
\begin{equation}
  \L(\phi, b) + \frac1{2\pi} Adb \Leftrightarrow
  \L(\psi, a) + \frac12\frac1{4\pi}ada + \frac1{2\pi}Ada + \frac1{4\pi}AdA .
\end{equation}
Using the time-reversed version Eq.~(\ref{fermion_fa}) and applying
charge conjugation, we find
\begin{equation}
   \L(\phi, a) + \frac1{2\pi} Ada \Leftrightarrow
  \L(\tilde\phi, A) ,
\end{equation}
which is the well known bosonic particle-vortex
duality~\cite{Peskin:1977kp,Dasgupta:1981zz}.  Although the duality is
well known, the success in deriving it gives one some confidence of
the correctness of the original seed duality.

A similar set of manipulations lead to the fermionic particle-vortex
duality, we start from the duality~(\ref{fermion_fa}), moving the
$AdA$ term to the right-hand side,
\begin{equation}
  \L(\Phi, A) \Leftrightarrow \L(\psi, a) - \frac12 \frac1{4\pi} ada
  + \frac1{2\pi}Ada - \frac1{4\pi}AdA .
\end{equation}
Now we gauge $A$ on both sides ($A\to b$), but add to the action
$-\frac1{4\pi}bdb+\frac1{2\pi}Adb$ before integrating over $b$,
\begin{equation}
  \L(\phi, b) - \frac1{4\pi}bdb + \frac1{2\pi}Adb
  \Leftrightarrow 
  \L(\psi, a) - \frac12 \frac1{4\pi} ada
  + \frac1{2\pi}bda - \frac2{4\pi}bdb + \frac1{2\pi}Adb .
\end{equation}
Using the time-reversed version of Eq.~(\ref{boson_fa}) one can
transform the left-hand side to a fermionic theory.  After moving the terms $AdA$ between sides, one finds at the end
\begin{equation}\label{careful_duality}
  \L(\tilde\psi, A) \Leftrightarrow
  \L(\psi, a) - \frac12 \frac1{4\pi} ada
  + \frac1{2\pi}bda - \frac2{4\pi}bdb + \frac1{2\pi}Adb
  - \frac12\frac1{4\pi} AdA .
\end{equation}
This is more accurate version of the fermionic particle-vortex duality
that would underlie the CF picture of the half-flled
Landau level.  The previous, na\"i{}ve version of the duality is
obtained by trying to integrate over $a$.  The saddle point equation
for this equation is
\begin{equation}
  db -2da + dA = 0,
\end{equation}
and if one sets $a=(b+A)/2$ (which in general violates the
quantization condition) one obtains the fermionic duality mentioned
above.  The fact that the right-hand side of
Eq.~(\ref{careful_duality}) is a time-reversal symmetric theory is not
obvious, but can be demonstrated~\cite{Seiberg:2016gmd}.

A whole web of dualities can be derived from the basic
boson--fermion duality in a manner similar to the one we have used
above.  A full account of these duality is beyond the scope of this
review.  Some remarkable insights obtained in this direction include a
self-dual theory of Ref.~\cite{Xu:2015lxa} and various descriptions of
a possible SO(5) quantum critical point~\cite{Wang:2017txt}.

\section{CONCLUSION}

We have presented arguments in favor of the Dirac nature of the
CF.  The Dirac CF provides a very simple
solution to a number of puzzles that have been plaguing the field
theory of the CF for a long time.  Regarded as a
low-energy effective theory, the Dirac CF theory,
modified to include the electric dipole moment of the CF, provides a description consistent with
low-energy constraints.

We have also reviewed various physical consequences of the Dirac
CF theory.  While most features of the HLR theory is
preserved by the Dirac CF theory, the latter leads to distinct
physical consequences, which are only started to be explored
experimentally~\cite{Pan:2017}.

\section*{DISCLOSURE STATEMENT}

The author is not aware of any affiliations, membership, funding, or
financial holdings that might be perceived as affecting the
objectivity of this review.

\section*{ACKNOWLEDGMENTS} 

This work is supported, in part, by US Department Of Energy grant
No.\ DE-FG02-13ER41958, the Army Research Office--Multidisciplinary
University Research Initiative grant No.\ 63834-PH-MUR, and a Simons
Investigator Grant from the Simons Foundation.  Additional support was
provided by the Chicago Materials Research Science \& Engineering
Center, which is funded by the National Science Foundation through
grant DMR-1420709.

\bibliography{arcmp_dcf-v3}
\bibliographystyle{ar-style4}

\end{document}